\documentstyle[epsfig,prl,aps,amsfonts,twocolumn]{revtex}
\begin{document}
\draft
\title{Percolation mechanism for Colossal Magnetoresistance}
\author{Paul J. M. Bastiaansen\cite{paul} and Hubert J. F. Knops}
\address{Institute for Theoretical Physics,
University of Nijmegen, Toernooiveld, 6525 ED Nijmegen, The Netherlands}
\maketitle

\begin{abstract}
We argue that colossal magnetoresistance is a critical phenomenon and
propose a mechanism to describe it. The mechanism relies on the
halfmetallic behavior of the materials showing colossal magnetoresistance,
and yields a correlated percolation model that, we argue, captures all
qualitative features of colossal magnetoresistance, above as well as below
the Curie temperature. The model only serves for revealing the
underlying mechanism of colossal magnetoresistance, and does not aim to
reproduce precise, numerical results.
\end{abstract}

\pacs{72.15.Gd, 64.60.Cn, 73.50.-h, 73.50.Jt}

\section{Introduction}

In recent years, much experimental effort has been devoted to
materials displaying Colossal Magnetoresistance (CMR), which is a strong
dependence of the resistance on the magnetic field as well as on
temperature. The studied materials are rare earth manganese
perovskites\cite{kusters,chahara,helmolt93,mccormack,helmolt94,jin,%
schiffer,gong,ju}, but recently the effect has been
shown\cite{shimakawa} to occur also in Tl$_2$Mn$_2$O$_7$, which has a
pyrochlore structure. These materials show a
ferromagnetic-to-paramagnetic transition at a certain temperature, the
Curie temperature. The resistance of the materials shows a peak at or
near this temperature, and falls off quite rapidly with higher and
lower temperatures. Switching on a magnetic field also lowers the
resistance. It is this behavior of the resistance as a function of
temperature and magnetic field that is called CMR.

The manganese perovskites are of a mixed valence type; if only Mn$^{3+}$
or Mn$^{4+}$ are present the material is insulating and
antiferromagnetic, and no CMR is observed. It is known for a long time
that the double exchange interaction between pairs of Mn$^{3+}$ and
Mn$^{4+}$ is responsible for the ferromagnetic and metallic properties
of the perovskites\cite{zener}. In this picture, a dependence of the
conductance on the spin direction of the charge carriers is already
present. More recently band structure calculations\cite{pickett}
using the local spin-density approximation indicated an effective
halfmetallic behavior, and in similar calculations adopting the
generalized gradient approximation the halfmetallic character fully
emerged\cite{peter}. The double exchange mechanism thus is responsible
for the halfmetallic behavior of these materials, be it that the valence
electrons are less localized than in the original model of
references~\onlinecite{zener}.

It is observed in all measurements of CMR that the peak in
the resistance occurs close to or at the Curie temperature~$T_c$. At
this temperature the spontaneous magnetization of the materials
vanishes; it is a critical point. The occurrence of the peak in the
resistance at or close to this temperature is a strong indication that
CMR itself is a critical phenomenon, that is, that the behavior of the 
resistance is intimately connected with the critical behavior of the
domains of magnetization. If this is true, typical band structure
calculations require, due to the infinite correlation length,
infeasible large system sizes to reproduce the effect.

The goal of this paper is to propose a mechanism for CMR. Relying on the
halfmetallic character of the materials displaying CMR, we will
introduce a model that, we believe, captures the basic features of the
phenomenon, but does not aim to reproduce the correct numerical
values of relevant temperatures, resistances and so on. Indeed, our
proposed model is much too simple as compared to the no doubt very
complicated processes that govern CMR. On the other hand, revealing the
basic mechanism that is responsible for the typical features of CMR is
the first step towards a satisfying understanding of the phenomenon.

Moreover, in our proposed mechanism, CMR is a critical phenomenon. The 
concept of universality in critical phenomena then assures that the
universal features of the phenomenon, such as qualitative behavior,
scaling functions, and critical exponents, will be reproduced
correctly, as they do not depend on the precise definition of the
model, but only on general features like symmetries and dimensionality.
Other quantities, like the precise location and height of the peak
in the resistance, do depend on the precise definition of the model,
and will, regarded the extreme simplicity of our model, be reproduced
incorrectly. The universal properties, however, can be put to strict
experimental tests to decide upon the validity of the proposed
mechanism.

The mechanism we have in mind is that of correlated percolation, and the
typical model that brings in the resistance is the correlated resistor
network. The model is introduced in section~\ref{model}. In
section~\ref{percolation} we will shortly explain the concepts of
percolation and resistor networks, and discuss the phase diagram of our
model. In this section, we will argue that the qualitative features of
CMR that are exhibited in experiments are captured by the model. In
section~\ref{resistance} we present Monte Carlo calculations on the
resistor network that make our results more explicit. In
section~\ref{discussion} we will discuss several complications that
arise when the model is extended to become a more sophisticated
explanation of CMR. We end with a conclusion.

\section{A model for colossal magnetoresistance}
\label{model}

Materials displaying CMR turn out to be
halfmetallic\cite{zener,pickett,peter}, that is, the band structure of
the electrons depends on the relative orientation of the spin with
respect to the magnetization of the material. Electrons with spin
parallel to the local magnetization are conducting, electrons with spin
antiparallel are insulating. The conductance of the material thus
depends on the structure of the domains with different directions of
magnetization. As suggested in Ref.~\onlinecite{ju}, this explains the
observation that the resistance increases with temperature, as more
domain walls occur with higher temperatures, thereby hampering the
percolation of charge carriers. It is less clear that this effect also
explains the decrease in resistance above $T_c$, but we will see that it
does, so there is no need to invoke another mechanism (e.g., magnetic
polarons) to explain the behavior above $T_c$.

The simplest conceivable model that captures such a mechanism is a lattice
model, where on every lattice point a vector $\bf m$ is defined, which
represents locally the direction of magnetization in the material. Due
to the extended wave function of the electrons, such a lattice point
does not have to be identified with one atom; the lattice can be defined
on a more coarse grained level as well. The direction of magnetization
$\bf m$ must in some way be governed by an energetic interaction between
the different domains, in such a way that for temperatures above the
Curie temperature $T_c$ the average magnetization vanishes, whereas
below the Curie temperature there is a finite magnetization. The
allowed orientations of $\bf m$ in each domain must follow the magnetic
anisotropy of the material, which is in the case of the CMR materials
a perovskite-like symmetry.

It is not our goal to introduce a precise, quantitative model that
describes the perovskites. We want to have a simple model that only
reveals the basic mechanism of CMR. Let us therefore, for simplicity,
assume that $\bf m$ can only point in two different directions, up and
down. We assume, then, that $\bf m$ obeys an Ising symmetry, so let us
replace $\bf m$ by the more familiar notation of a scalar spin
$S\in\{+1,-1\}$. The interaction between the different domains is also
defined in the most simple manner, that is, there exists a `nearest
neighbor' coupling $K$ between the spins. In addition, we allow for an
external magnetic field $h$. The Hamiltonian governing the statistics
of the model is in this case the familiar Ising Hamiltonian
\begin{equation}
  H = -K \sum_{\langle ij\rangle} S_iS_j - h\sum_j S_j,
  \label{hamiltonian}
\end{equation}
where $\langle ij\rangle$ denotes a summation over nearest neighbor
pairs only.

The resistance is brought into the model following the half-metallic
character of the CMR materials. Resistances are defined independently for 
charge carriers with spin up and down. Charge carriers with spin up have
a low resistance in domains characterized by $S=+1$ and a high
resistance in domains with $S=-1$. For these spin-up carriers we define
a unit resistance on bonds between two sites having $S=+1$ and an
infinite resistance on $(+-)$ and $(--)$ bonds. The infinite resistance
on a $(--)$ bond reflects the insulating character and is a
simplification, as these domains will have a large but finite
resistance. The infinite resistance on $(+-)$ bonds means that our
model does not allow for spin-flip processes, and this is no
doubt a simplification as well.

For the charge carriers with spin down a similar assignment of
resistances is made. These assignments yield an expectation value of the
overall resistance of the material independently for up and down
carriers. We denote the inverse overall resistance, i.e., the
conductance, by $\langle\sigma_+\rangle$ and $\langle\sigma_-\rangle$
respectively. The total overall conductance then simply is
\[
  \langle\sigma\rangle = \langle\sigma_+\rangle +
						 \langle\sigma_-\rangle.
\]
The angular brackets denote Ising expectation values.

\begin{figure}
  \begin{center}
  \epsfig{file=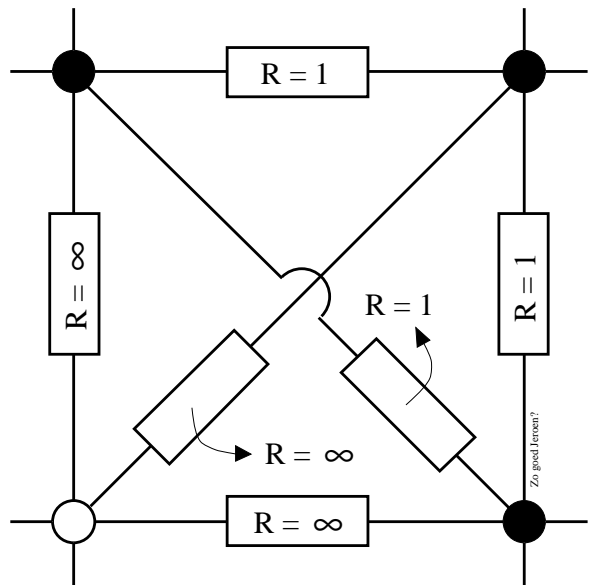,width=0.9\hsize}
  \end{center}
  \caption{The assignment of resistors between neighbor and
  next-neighbor sites for the charge carriers with spin up. The small
  circles denote the lattice sites; black sites have a magnetization
  $S=+1$, white sites have $S=-1$. If both sites have $S=+1$ a unit
  resistance is put on the bond between them, otherwise the bond
  receives an infinite resistance. A similar assignment is made for the
  charge carriers with spin down.}
  \label{resistor}
\end{figure}

The assignment of resistances for the spin up carriers is shown in
figure~\ref{resistor}, where sites with $S=+1$ are depicted black, and
sites with $S=-1$ are white. In the figure resistances are defined
between nearest as well as next-nearest neighbor sites; the reason
for this is explained below.

The conductance of the model is calculated in the usual way: let
$\Gamma$ be a configuration of magnetizations $S_j$ on the lattice
sites. The conductance of this configuration can be calculated from the
assignment of resistances, and is denoted by $\sigma(\Gamma)$. The
expectation value of the overall conductance then is obtained by a
weighted summation over all configurations in the usual way,
\[
  \langle\sigma\rangle = \frac1Z \sum_\Gamma 
	 \sigma(\Gamma)\;\exp\big(-\beta H(\Gamma)\big),
\]
where $\beta=1/kT$, $Z$ is the partition function of the Ising model and
$H(\Gamma)$ is the Hamiltonian of equation~(\ref{hamiltonian}). We will
now see what the predictions of this model are for the expectation
value of the overall conductance of the lattice.

\section{The percolation phase diagram}
\label{percolation}

The overall conductance of the lattice is defined as follows: consider a
square lattice of $L\times L$ sites with resistances on bonds defined as
in figure~\ref{resistor}. Keep the lower row of the lattice at a fixed
potential $V=0$ and the upper row at $V=1$. The conductance of the
lattice is then equal to the expectation value of the current.

Conductance over the lattice is possible (e.g., non-zero) only if
there exists a path from border to border exclusively over resistances
$R=1$, that is, if at least one of the two spin directions is
percolating\cite{essam,kirkpatrick}. The qualitative features of the
model thus can be obtained by considering the percolation phase diagram
of the Ising model. In percolation, one defines clusters by putting
bonds between neighbor spins $S=+1$ or $S=-1$. Let us consider
percolation for sites $S=+1$; percolation for $S=-1$ is of course just
the same but with the Ising magnetic field $h$ replaced by $-h$.  In
our case, a bond is placed between each pair of nearest and
next-nearest neighbor sites having $S=+1$. These bonds make up
clusters, and the percolation order parameter $P$ is the density of
sites in the percolating cluster. If there is percolation then $P$ is
finite, if there is no percolation, $P=0$.

\begin{figure}
  \begin{center}
  \epsfig{file=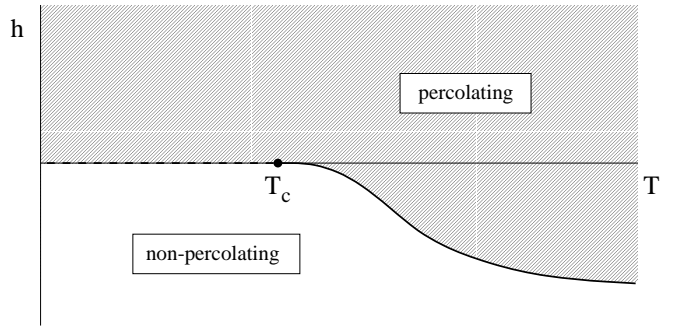,width=5cm,angle=-90}
  \end{center}
  \caption{The percolation phase diagram for the Ising model for
  percolation of the up spins. Percolation is possible between
  nearest and next-nearest neighbor up spins. $T$ is temperature and
  $h$ is the magnetic field. The thick, solid line is a critical
  percolation line. The order parameter (the fraction of sites present
  in the largest
  cluster) is finite in the shaded area and goes continuously to zero
  when approaching the critical line. The critical line merges smoothly
  with the $T$-axis at $T_c$, which is a tricritical point for
  percolation. The dashed line is a first order transition for
  percolation, meaning that the order parameter jumps from a finite
  value to zero on this line.}
  \label{phasediag}
\end{figure}

Our case is called correlated percolation as the distribution of
percolating ($S=+1$) and non-percolating ($S=-1$) sites is a correlated
one. When this distribution is that of the Ising model, the phase
diagram is known, and this is one of the reasons of choosing the Ising
model as an example of our model: many exact results, also for
percolation, are known for this model. Its percolation phase diagram is
known\cite{stella} when percolating bonds are placed between each
nearest neighbor pair of sites having $S=+1$. In another
paper\cite{wij}, we derive the phase diagram when next-nearest neighbor
percolating bonds are defined as well. It is this phase diagram we need
for our explanation; it is shown in figure~\ref{phasediag}.

In experiments, CMR occurs in thin
films\cite{chahara,helmolt93,mccormack,helmolt94} or even in bulk
samples\cite{kusters,helmolt94,ju}. Our model thus should be pseudo
three-dimensional or fully three-dimensional. This is the reason that
in our two-dimensional model we defined resistances on next-nearest
neighbor bonds as well. These `crossing' bonds make the bond-graph {\it
non-planar}, and this is the simplest way to mimic pseudo
three-dimensional behavior. By defining these crossing bonds we allow
for more possible paths between percolating sites, and the same is
happening in a system consisting of several layers.

Inclusion of these crossing bonds is crucial, because without the
crossing bonds there is no percolation\cite{stella} above the Curie
temperature. At first it seems therefore that the percolation mechanism
cannot explain the decreasing resistance above $T_c$. Inclusion of the
crossing bonds, however, makes the region above $T_c$ percolating, and
this is necessary for our explanation.

\begin{figure}
  \begin{center}
  \epsfig{file=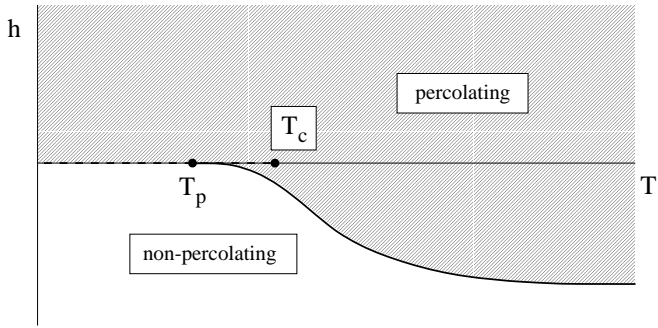,width=5cm,angle=-90}
  \end{center}
  \caption{The phase diagram for a lattice model where the percolation
  point $T_p$ and the Curie point $T_c$ do not coincide. In this case
  the percolation temperature is always lower than the Curie
  temperature. The thick solid line is a critical percolation line in
  the universality class of random percolation and the same is expected
  for the endpoint of the line. The dashed line remains a first order
  transition line.}
  \label{phasediag-ext}
\end{figure}

The effect of even more possible paths or of bonds with an even larger
percolation range is, however, quite small. Even in a fully
three-dimensional model, where there are many more possible paths for a
cluster to follow, the percolation point lies only a few percent below
the Curie temperature\cite{muller}. We expect, therefore, our
conclusions also to hold in the case of a different lattice model or a
different dimensionality. The percolation point may lie somewhat below
$T_c$ but will be close to it. The phase diagram in that case is shown
in figure~\ref{phasediag-ext}.

We choose to follow figure~\ref{phasediag} for the explanation of our
mechanism and the comparison with experiments. The qualitative
comparison, however, holds for the other phase diagram of 
figure~\ref{phasediag-ext} as well.

For $T\rightarrow\infty$ or $K\rightarrow0$ the probability
distribution becomes a random one and the type of percolation is random
percolation. The thick line merging smoothly with the $T$-axis is in the
universality class of random percolation: on this line the order
parameter vanishes algebraically with the same exponent as that for
random percolation. The Curie point $T_c$ is the critical point of the
Ising model, where there is a ferromagnetic-to-paramagnetic transition.
For percolation, this point is a tricritical percolation point. The
dashed line is a first order transition for the magnetization, and for
percolation as well; the order parameter $P$ jumps from a finite value
to zero over this line.

In principle, the percolation phase diagram tells nothing about the value 
of the conductance, only whether it is zero or non-zero. Nevertheless
we may well assume that the closer one moves to the percolation
threshold, the lower the conductance will be. From this, we can extract
the qualitative behavior of the overall resistance from the phase
diagram and draw a comparison with the experimental properties of CMR.
To check the predictions of the phase diagram, we performed Monte Carlo
calculations on the Ising resistor network. The results of these
calculations confirm the predictions of the phase diagram, and are
discussed in the next section.

Our conclusion is that the qualitative features of CMR are all present
in the phase diagram of correlated percolation. Consider the following
features:

1) The peak in the resistance at the Curie temperature $T_c$. According
to the phase diagram, the only point on the $T$-axis where no
percolation occurs for both spin directions is the Curie point $T_c$, so
there is a peak in the resistance at $T_c$. (In fact, in our model this
peak is infinitely high due to the choice $R=\infty$ on the insulating
bonds.) All other points on the $T$-axis are percolating and hence show
a better conductance.

2) The asymmetric shape of the peak in the resistance as a function of
$T$. This shape is clearly present\cite{kusters,schiffer,gong,shimakawa}
in experiments. In the phase diagram, moving away from $T_c$ is moving
away from the percolation point and hence decreasing the resistance. Due 
to the presence of the critical percolation line that merges 
smoothly\cite{stella} with the $T$-axis, we expect this decrease to be slower 
above $T_c$, resulting in an asymmetric peak in the resistance. This
expectation is confirmed using a Monte Carlo calculation, described in
the next section.

3) The peak in the resistance as a function of magnetic field. In every 
experiment, the resistance peaks at zero field, a
feature that is clearly present in the phase diagram. Moving away from
the zero field axis means stimulating the percolation of one of the spin
directions and hence decreasing the resistance. Note that our model does
not give a correct numerical value of the magnetoresistance ratio
$\Delta R/R$, where $\Delta R$ is the difference in resistance with and
without magnetic field. What does follow from the model, however, is the
qualitative result that this ratio is largest at $T_c$.

4) A particular observation in experiments is the shape of the peak in
the resistance as a function of the magnetic field. As can be seen, e.g.,
from Fig.~1 in Ref.~\onlinecite{mccormack}, Fig.~5 in
Ref.~\onlinecite{helmolt94}, and Fig.~5 in Ref.~\onlinecite{jin},
plots of the resistance versus the field display a cusp at temperatures
below $T_c$ but are smooth at higher temperatures. This remarkable
feature follows directly from the
percolation phase diagram: below $T_c$ there is a first order transition
line. Moving away from this line in both directions of the field
immediately decreases the resistance, meaning that a plot of the
resistance versus the field shows a cusp, the peak being strongest at
$T_c$. Above $T_c$, on the other hand, there is no transition line at
zero field, and hence the behavior of the resistance as a function of
the magnetic field must be smooth.

5) It is observed\cite{kusters,helmolt94,schiffer,shimakawa} in CMR
that the peak in the resistance shifts to higher temperatures when a
magnetic field is switched on. This directly follows from the presence
of the percolation line in the phase diagram at temperatures above
$T_c$. Switching on the magnetic field above $T_c$ one `feels' the
vicinity of this line, whereas below $T_c$ the nearest percolation
point is $T_c$ itself, which is further away. Hence the resistance
decreases more strongly as a function of the field below $T_c$.

This comparison shows that the basic features of CMR that are displayed
in all experiments, are directly explained by our simple percolation
mechanism. One single and simple mechanism accounts for its features,
above as well as below the Curie temperature.

\section{Resistance calculations}
\label{resistance}

To make some predictions of our model, as explained in the previous
section, more explicit, we performed Monte Carlo calculations on the
Ising correlated resistor network. The model we used is that described
in section~\ref{model} with the Hamiltonian of
equation~(\ref{hamiltonian}) and the assignment of resistors of
figure~\ref{resistor}.

We performed Monte Carlo simulations of the expectation value of the
conductance for several lattice sizes. We used the 
Wolff-algorithm\cite{wolff} for the Monte Carlo part, and the multigrid 
method of Edwards, Goodman and Sokal\cite{sokal}, based on the standard
code {\sc amg1r4}\cite{amg1r4}, to calculate the conductance of a spin
configuration. The latter code is slightly changed, in order to cope
with resistors on next-nearest neighbor bonds.

\begin{figure}
  \begin{center}
  \epsfig{file=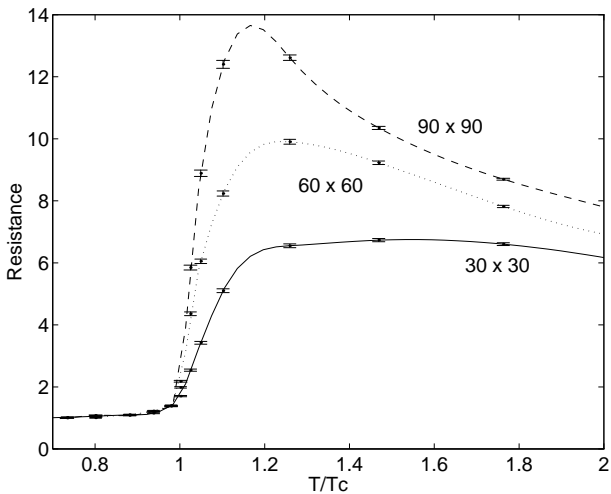,width=0.5\textwidth}
  \end{center}
  \caption{Resistance as a function of temperature for different
  system sizes at zero field. The system sizes are indicated in the
  plot, the resistance is scaled such that the minimal resistance (all
  bonds having $R=1$) is unity. The lines are guides to the eye.
  Resistors are defined as in figure~\ref{resistor}.
  For larger systems, the peak shifts to $T_c$, becomes infinitely high
  but remains asymmetric. For $T<T_c$ there is a rapid convergence to the
  minimal (unit) resistance.}
  \label{phaseline}
\end{figure}

The interest in the first place is the behavior of the resistance for
different temperatures. We performed calculations on different system
sizes; the result is shown in figure~\ref{phaseline}. The peak in the
resistance at $T_c$ becomes, for an infinite system, infinitely high, a
result already following from the percolation phase diagram. The
calculations confirm also the asymmetric shape of the peak in the
resistance.

Secondly, the interest is in the critical exponents of the correlated
resistor network. The interested reader is referred to
reference~\onlinecite{wij} for a more detailed account. In this
reference, we describe our calculations of the conductance exponent $t$
of the correlated resistor network. This exponent governs the algebraic
decay of the conductance upon approaching a percolation threshold. Two
different exponents play a role in the phase diagram of
figure~\ref{phasediag}. The first is the exponent of the critical
percolation line. It is defined as
\[
   \sigma(h) \sim |h-h_c|^t,
\]
where $\sigma(h)$ is the conductance at fixed $T>T_c$, and $h_c$ is the
critical value of the magnetic field. The critical percolation line is
in the universality class of random percolation, so the value of the
exponent $t$ on this line must be that of the random resistor network.
The random resistor network is a long standing unsolved problem in
statistical physics, but good numerical results of the exponent $t$ are
known. The best estimate\cite{normand} in two dimensions known to us is
$t=1.299\pm0.002$.

The second exponent involved is that governing the vanishing conductance
at the Ising critical point. For percolation, this point is in a
different universality class, and therefore the conductance exponent
also differs from the random one. We calculated this exponent from the
finite size behavior of the conductance, as explained in
reference~\onlinecite{wij}. Because the Ising critical point is a
tricritical point for percolation, there are two non-equivalent
directions of approaching the critical point with two different
exponents involved.  For the temperature direction, the exponent $t$ is
defined as
\begin{equation}
   \sigma(T) \sim |T-T_c|^t \quad\text{at $h=0$},
   \label{tricrit}
\end{equation}
yielding a value $t = 0.2000\pm 0.0007$. For the other, field-like,
direction, the exponent is defined as
\[
   \sigma(h) \sim |h|^t \quad\text{for $T=T_c$}.
\]
In this case, the exponent $t$ has the value $t=0.1067\pm0.0004$.

As will be explained in the next section, the numerical values of these
exponents are of no direct interest, as the Ising model is not the
correct lattice model for describing the CMR-materials. What is of
interest, however, is the difference between the exponent of the
critical line (random resistor network) and those of the Ising critical
point (correlated resistor network). These exponents differ
considerably. Even when there turns out to be no tricritical point in
the correct phase diagram of the CMR-materials, as in
figure~\ref{phasediag-ext}, the model will, in a sense, be close to the
tricritical point, which means that there will be an observable
crossover from the random exponent to the value of the tricritical
exponents.

\section{Discussion}
\label{discussion}

We chose the Ising model to serve as an example for our mechanism for
CMR, because of its simplicity and for the fact that many exact results
are known for this model. There are, however, several features of our
model that render it unrealistic for a more sophisticated explanation
of CMR. The first is of course that the Ising model is not the
appropriate lattice model, as the Ising variables $S_j$ do not have the
symmetry of the CMR-materials. The latter are mostly perovskites, so
the appropriate lattice model should have a magnetization vector $\bf
m_j$ at each site $j$ that follows the symmetry of the perovskites.

A further shortcoming of our model is the infinite resistance for
spin-up charge carriers on $(+-)$ and $(--)$ bonds. In principle, we
should allow for a finite (but large) resistance on these bonds. The
effect of this can be compared with the ferromagnetic-to-paramagnetic
phase transition in the Ising model: if we switch on a small magnetic
field, the true phase transition disappears, but the magnetization,
specific heat and so on still show signs of the vicinity of the
critical point. The same will happen when introducing these large but
finite resistances.  The actual value of this resistance displays the
possibility of a spin-flip mechanism of the charge carriers. The
importance of this effect is, we believe, quite small, such that the
true phase transition is smeared out but the behavior of the overall
resistance still shows signs of the vicinity of the percolation
transition.

The finite height of the peak in the measurements can simply be
incorporated by a large but finite `background' resistance $R_\infty$
parallel to the net resistance of the correlated resistor network,
representing the resistance of the `non-conducting' bonds. The total
resistance as a function of temperature $T$, following from
equation~(\ref{tricrit}), then becomes
\begin{equation}
   R(T) = \frac{R_\infty}{1 + a|T-T_p|^t},
   \label{shunt}
\end{equation}
where $T_p$ is the temperature where the resistance peaks, so $T_p$ is
equal to or slightly lower than the Curie temperature $T_c$. The
amplitude $a$ is non-universal and has in general different values
above and below $T_p$.

Experiments so far have not been set up to measure critical exponents.
For the exponent $t$ of equation~(\ref{shunt}), a rough analysis of the
published data seems to point to a value between $1.5$ and $2.5$. It
would be worthwhile to have more accurate experimental data on the
exponents. 

On the theoretical side, the interest is in the critical exponents of
the correlated resistor network for the spin model corresponding to the
magnetic anisotropy of the perovskites. Calculating these exponents is
considerably more elaborate than in the Ising case, as even the
percolation phase diagram has, to our knowledge, never been studied for
models other than Ising.

\section{Conclusions}

We presented a correlated percolation mechanism for CMR and argued that
its observed effects can be explained by considering the materials
displaying CMR as a correlated resistor network. The introduced model
only sheds light on the underlying mechanism, but is certainly not the
appropriate one to reproduce accurate numerical results. For this, the
model has to be changed into a spin model displaying the right magnetic
anisotropy of the perovskites and has to be made pseudo or fully
three-dimensional. Nevertheless, we argue, the qualitative results of
the model will remain untouched.

\begin{acknowledgments}
We thank Rob de Groot and Peter de Boer for drawing our attention to
the problem of CMR and for enlightening discussions; Alan Sokal for
providing us with the code for the resistance calculations, which saved
us a lot of work; Bernard Nienhuis for discussions on correlated
percolation; and Erik Luijten for discussions on the Monte Carlo part.
\end{acknowledgments}

\end{document}